\documentclass[english,letterpaper,aps,prl,twocolumn,showpacs]{revtex4}
\usepackage[T1]{fontenc}
\usepackage[latin1]{inputenc}
\usepackage{graphicx}
\usepackage{amssymb}

\makeatletter

\makeatletter

\usepackage{epsfig}

\usepackage{dcolumn}

\usepackage{bm}

\usepackage{bbm}

\usepackage{wasysym}

\usepackage{marvosym}

\usepackage{times}

\newlength{\textwidthm}

\makeatother

\usepackage{babel}
\makeatother

\begin{document}

\title{Localized Magnetic States in Graphene}

\author{Bruno Uchoa$^{1}$, Valeri N. Kotov$^{1}$, N.~M.~R. Peres$^{2}$,
and A.~H. Castro Neto$^{1}$}

\affiliation{$^{1}$Department of Physics, Boston University, 590 Commonwealth
Avenue, Boston, MA 02215, USA}

\affiliation{$^{2}$Centro de F\'{\i}sica e Departamento de F\'{\i}sica, Universidade
do Minho, P-4710-057, Braga, Portugal}

\date{\today}

\begin{abstract}
We examine the conditions necessary for the presence of localized
magnetic moments on adatoms with inner shell electrons in graphene.
We show that the low density of states at the Dirac point, and the
anomalous broadening of the adatom electronic level, lead to the formation
of magnetic moments for arbitrarily small local charging energy. As
a result, we obtain an anomalous scaling of the boundary separating
magnetic and non-magnetic states. We show that, unlike any other material,
the formation of magnetic moments can be controlled by an electric
field effect. 
\end{abstract}


\maketitle
Graphene, a two-dimensional (2D) allotrope of carbon, has singular
spectroscopic and transport properties \cite{novo1,novo3,kim3,geim}
due to its unusual electronic excitations described in terms of massless,
chiral, {}``relativistic'' Dirac fermions \cite{Antonio}. Besides
being a possible test bed for relativistic quantum field theory \cite{pw},
graphene has a great technological potential due to its structural
robustness, allowing extreme miniaturization \cite{pono}, and a
flexible electronic structure that can be controlled by an applied
perpendicular electric field \cite{edu}.

In this paper we show that graphene has also potentiality for \textit{spintronics},
that is, independent control of the charge and the spin of the charge
carriers \cite{wolf}. Unlike diluted magnetically semiconductors
(DMS) \cite{allan} where the location of the magnetic ions is random
and hence unpredictable, adatoms can be positioned in graphene using
a scanning tunneling microscope (STM) \cite{eigler}. Furthermore,
as we are going to show, the magnetic properties of adatoms such as
size of the magnetic moment and Curie temperatures can be controlled
by an external electric field, an effect unparalleled in condensed
matter systems.

The basic model for the study of magnetic moment formation in metals
is the well-known Anderson impurity model \cite{anderson61}. In
this model an ion with inner shell electrons with energy $\epsilon_{0}$
hybridizes, via a hopping term of energy $V$, with a conduction sea
of electrons. While the conduction electrons are described by a Fermi
liquid with featureless, essentially constant, density of states (DOS),
the impurity ion is assumed to be strongly interacting. The Coulomb
energy required for double occupancy of an energy level in the ion
is given by $U$. Anderson showed that when $\epsilon_{0}$ is below
the Fermi energy, $\mu$, and the energy of the doubly occupied states,
$\epsilon_{0}+U$, is larger than $\mu$, a magnetic state is possible
if $U$ is sufficiently large and/or $V$ sufficiently small.

Here we apply the Anderson model to graphene and show that the energy
dependence of the DOS leads to anomalous broadening of the adatom
level and strongly favors the formation of local magnetic moments.
In particular we show that, unlike the case of ordinary metals, this
anomalous broadening allows the formation of magnetic states even
when $\epsilon_{0}$ is \textit{above} the Fermi energy at relatively
small $U$. We also find, in contrast with the usual metallic case,
an anomalous scaling of the magnetic boundary separating magnetic
and non-magnetic impurity states. Finally, we establish that the local
magnetic moments can be mastered by the application of an external
gate voltage, leading to a complete control of the magnetic properties
of adatoms in graphene.

\begin{figure}[b]

\begin{centering}
\includegraphics[clip,width=8.6cm]{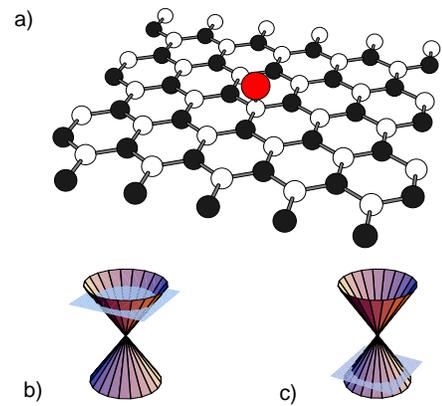} 

\par\end{centering}

\caption{(color on line) (a) Honeycomb lattice with an impurity atom. Black:
sublattice A; White: sublattice B. Intersection of the Dirac cone
spectrum, $\epsilon_{\pm}({\bf k})=\pm v_{F}k$, with the localized
level spectrum, $E_{f}({\bf k})=\epsilon_{0}$: b) $\epsilon_{0}>0$;
(c) $\epsilon_{0}<0$.}

\label{Fig_honey_boron} 
\end{figure}

We consider an impurity atom adsorbed on the surface of the graphene
sheet, on top of a carbon (see Fig. \ref{Fig_honey_boron}). The tight-binding
Hamiltonian of the electrons in graphene is \begin{eqnarray}
H_{TB} & = & -t\sum_{\sigma}\sum_{\langle i,j\rangle}\left[a_{\sigma}^{\dagger}(\mathbf{R}_{i})b_{\sigma}(\mathbf{R}_{j})+H.c.\right],\label{HTB}\end{eqnarray}
 where $a_{\sigma}({\bf R}_{i})$ ($b_{\sigma}({\bf R}_{i})$) annihilates
and electron with spin $\sigma=\uparrow,\downarrow$ on sublattice
$A$ ($B$) at position $\mathbf{R}_{i}$, $\langle i,j\rangle$ stands
for summation over nearest neighbors, and $t$ ($\approx2.7$ eV)
is the nearest neighbor hopping energy. In momentum space, we have
(we use units such that $\hbar=1$): \begin{eqnarray}
H_{TB} & = & -t\sum_{\mathbf{k},\sigma}\left[\phi(\mathbf{k})a_{\mathbf{k},\sigma}^{\dagger}b_{\mathbf{k},\sigma}+\phi^{*}(\mathbf{k})b_{\mathbf{k},\sigma}^{\dagger}a_{\mathbf{k},\sigma}\right]\,\label{HTB2}\end{eqnarray}
 where $\phi(\mathbf{k})=\sum_{\vec{\delta}}\mbox{e}^{i\mathbf{k}\cdot\vec{\delta}}$,
with $\vec{\delta}_{1}=a(\hat{x}/2+\sqrt{3}/2\hat{y}),\,\vec{\delta}_{2}=a(\hat{x}/2-\sqrt{3}/2\hat{y})$
and $\vec{\delta}_{3}=-a\hat{x}$ are the nearest neighbor vectors.
Diagonalization of the Hamiltonian (\ref{HTB2}) generates two bands,
$\epsilon_{\pm}(\mathbf{k})=\pm t|\phi(\mathbf{k})|$, which can be
linearized around the Dirac points $\mathbf{K}$ at the corners of
the Brillouin zone: $\epsilon_{\pm}(\mathbf{K}+\mathbf{q})\approx\pm v_{F}|\mathbf{q}|$,
where $v_{F}=3ta/2$ ($\approx10^{6}$ m/s) is the Fermi velocity
of the Dirac electrons.

The hybridization with the localized orbital of the impurity atom
on a given site, say, on sublattice $B$, is given by: $H_{V}=V\sum_{\sigma}[f_{\sigma}^{\dagger}b_{\sigma}(0)+H.c.]\,,$
where $f_{\sigma}$ ($f_{\sigma}^{\dagger}$) annihilates (creates)
and electron with spin $\sigma=\uparrow,\downarrow$ at the impurity.
In momentum space we have: \begin{equation}
H_{V}=(V/\sqrt{N_{b}})\sum_{\mathbf{p},\sigma}(f_{\sigma}^{\dagger}b_{\mathbf{p},\sigma}+b_{\mathbf{p},\sigma}^{\dagger}f_{\sigma})\,,\end{equation}
 where $N_{b}$ is the number of sites on sublattice $B$ contained
in the expanded unit cell of graphene with the impurity.

The Hamiltonian of the localized orbital is described by a single
level, $H_{f}=\epsilon_{0}\sum_{\sigma}f_{\sigma}^{\dagger}f_{\sigma}$.
The electronic correlations in the inner shell states can be described
by a Hubbard-like term: $H_{U}=Uf_{\uparrow}^{\dagger}f_{\uparrow}f_{\downarrow}^{\dagger}f_{\downarrow}$.
Following Anderson, we use a mean-field decoupling of the interaction,
$H_{U}\rightarrow\sum_{\sigma}Un_{-\sigma}f_{\sigma}^{\dagger}f_{\sigma}-Un_{\uparrow}n_{\downarrow}$,
where $n_{\sigma}=\langle f_{\sigma}^{\dagger}f_{\sigma}\rangle$
is the occupation for each of the two spin states. The Hubbard term
can be absorbed into the definition of the local impurity energy,
$H_{f}=\sum_{\sigma}\epsilon_{\sigma}f_{\sigma}^{\dagger}f_{\sigma}\,,$
where $\epsilon_{\sigma}=\epsilon_{0}+Un_{-\sigma}$ is the energy
of the localized electrons in a given spin state in the presence of
a charging energy $U$.

The formation of a magnetic moment is determined by the occupation
of the two spin states at the impurity, $n_{\sigma}$. A localized
moment forms whenever $n_{\uparrow}\neq n_{\downarrow}$ The determination
of $n_{\sigma}$ requires the self-consistent calculation of the density
of states at the impurity level, $\rho_{ff}(\omega)$, which incorporates
the broadening of the impurity level due to hybridization with the
bath of electrons in graphene. The occupation of the impurity level
is given by: \begin{equation}
n_{\sigma}=\int_{-\infty}^{\mu}d\omega\,\rho_{ff,\sigma}(\omega)\,.\label{MFh2}\end{equation}

The Green's function of $f$-electrons is: $G_{ff,\sigma}(t)=-i\langle T\left[f_{\sigma}(t)f_{\sigma}^{\dagger}(0)\right]\rangle$,
and its retarded part can be written as: \begin{equation}
G_{ff,\sigma}^{R}(\omega)=\left[\omega-\epsilon_{\sigma}-\Sigma_{ff}^{R}(\omega)+i0^{+}\right]^{-1},\label{G_ff}\end{equation}
 where \begin{equation}
\Sigma_{ff}^{R}(\omega)=(V^{2}/N_{b})\sum_{\mathbf{p}}G_{bb,\sigma}^{0\, R}(\mathbf{p},\omega)\,,\label{G_bar}\end{equation}
 is the self-energy of the $f$-electrons, which is defined in terms
of the non-interacting Green's function of the graphene electrons
in a given sublattice, $G_{bb,\sigma}^{0}(\mathbf{p},t)=-i\langle T\left[b_{\sigma\mathbf{p}}(t)b_{\sigma\mathbf{p}}^{\dagger}(0)\right]\rangle_{0}$:
\begin{equation}
G_{bb,\sigma}^{0\, R}(\mathbf{p},\omega)=\omega/(\omega^{2}-v_{F}^{2}|\mathbf{p}|^{2}+i0^{+}\mbox{sign}(\omega))\,.\quad\label{Gbb}\end{equation}
 In this case (\ref{G_bar}) becomes: \begin{equation}
\Sigma_{ff}^{R}(\omega)=-V^{2}\frac{\omega}{D^{2}}\ln\!\left(\frac{\vert\omega^{2}-D^{2}\vert}{\omega^{2}}\right)-iV^{2}\frac{\pi\vert\omega\vert}{D^{2}}\theta(D-\vert\omega\vert)\,,\label{G_bar_R}\end{equation}
 where $D$ is a high-energy cut-off of the order of the graphene
bandwidth ($D\approx7$ eV). We choose the cut-off $D$ using the
Debye prescription, i.e., conservation of the number of states in
the Brillouin zone after linearization of the spectrum around the
$K$ point. We assume $|\mu|\ll D$, where band effects
related to the exact definition of the cut-off are not important.

The real part of $\Sigma_{ff}^{R}(\omega)$ defines the quasiparticle
residue $Z^{-1}(\omega)=1+(V^{2}/D^{2})\ln\left(|D^{2}-\omega^{2}|/\omega^{2}\right)$
of the $f$-electrons, while the imaginary part gives the broadening
of the localized level due to the hybridization. As expected, the
anomalous character of the problem is explicitly manifested in the
linear dependence of the broadening with the energy, which is proportional
to the electronic DOS in graphene. Furthermore, notice that $Z(\omega)$
vanishes at $\omega\to0$. Replacing Eq. (\ref{G_bar_R}) into Eq.
(\ref{G_ff}) gives the density of states of the localized level,
$\rho_{ff,\sigma}(\omega)=-1/\pi\mbox{Im}\, G_{ff,\sigma}^{R}(\omega)$:
\begin{equation}
\rho_{ff,\sigma}(\omega)=\frac{1}{\pi}\frac{\Delta\vert\omega\vert\theta(D-\vert\omega\vert)}{[Z^{-1}(\omega)\ \omega-\epsilon_{\sigma}]^{2}+\Delta^{2}\omega^{2}}\label{rhoff}\end{equation}
 where $\Delta=\pi V^{2}/D^{2}$ is the dimensionless hybridization.

Notice that, unlike the case of impurities in metals, the impurity
density of states is not a simple lorentzian. The impurity DOS, (\ref{rhoff}),
is peaked around the quasiparticle pole at $\epsilon_{\sigma}=0$
and $\omega\to0$. We can expand $Z(\omega)$ around the singularity
at $\omega_{0}Z^{-1}(\omega_{0})\sim\epsilon_{\sigma}$ for $\omega_{0}\to0$,
where we may approximate $Z(\omega_{0})\sim Z(\epsilon_{\sigma})$
except for a double-logarithmic corrections that can be safely ignored\cite{note}.
The anomalous broadening gives rise to a logarithmic divergence in
the ultraviolet when the DOS of the level is integrated in (\ref{MFh2}),
\begin{equation}
n_{\sigma}=\frac{Z_{\sigma}^{-1}}{(Z_{\sigma}^{-2}+\Delta^{2})}\!\!\left[\theta(\mu_\sigma-\epsilon_{\sigma})+\frac{1}{\pi}\mbox{arctan}\!\left(\frac{|\mu|\Delta }{\epsilon_{\sigma} -\mu_\sigma}\right)+\Theta_{\sigma}\right]\label{MF}\end{equation}
 where $Z(\epsilon_{\sigma})\equiv Z_{\sigma}$ and $\mu_\sigma \equiv \mu Z_\sigma^{-1}$. The term
$\Theta_{\sigma}$ contains the contribution coming from the cut-off
regularization: \begin{equation}
\Theta_{\sigma}\!=\!Z_{\sigma}\frac{\Delta}{\pi}\ln\!\left[\frac{W_{\sigma}(E_{\sigma})^{\gamma}}{(\epsilon_{\sigma})^{1+\gamma}}\right]-\frac{1}{\pi}\mbox{arctan}\!\left(\frac{\Delta D}{DZ_{\sigma}^{-1}+\epsilon_{\sigma}}\right),\label{Theta}\end{equation}
 where $\gamma=\mbox{sign}(\mu)$, and \begin{eqnarray}
E_{\sigma} & = & \sqrt{(\epsilon_{\sigma}-\mu Z_{\sigma}^{-1})^{2}+\mu^{2}\Delta^{2}}\label{E}\\
W_{\sigma} & = & \sqrt{(DZ_{\sigma}^{-1}+\epsilon_{\sigma})^{2}+D^{2}\Delta^{2}}\,.\label{W}\end{eqnarray}

\begin{figure}[btf]

\begin{centering}
\includegraphics[clip,width=8.6cm]{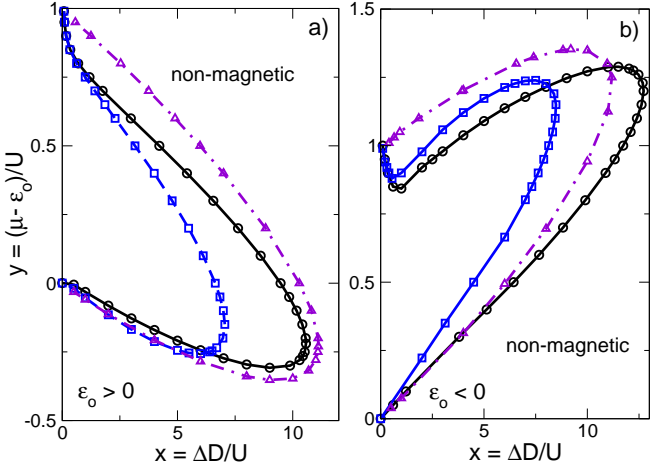} 
\par\end{centering}

\caption{Boundary between magnetic and non-magnetic impurity states in the
scaling variables $x$ and $y$ for $\epsilon_{0}>0$ (a) and $\epsilon_{0}<0$
(b). Circles: $|\epsilon_{0}|/D=0.029$, $V/D=0.14$; Squares: $\epsilon_{0}/D=0.043$
and $V/D=0.14$; Triangles:$|\epsilon_{0}|/D=0.029$, $V/D=0.03$.
The upturn close to $y=1$ and $x\to0$ on panel b) is not visible
in this scale when $V$ is very small (triangles). See details in
the text. }

\end{figure}

In Fig.~2 we show the boundary between magnetic and non-magnetic
impurity states as a function of the parameters $x=D\Delta/U$ and
$y=(\mu-\epsilon_{0})/U$. Notice that the magnetic boundary is not
symmetric between the cases where the impurity is above ($\epsilon_{0}>0$)
or below ($\epsilon_{0}<0)$ the Dirac point. Moreover, unlike the
metallic problem\cite{anderson61} the boundary is not symmetric
around $y=0.5$. This reflects the particle-hole symmetry breaking
due to the presence of the localized level. In the case where $\epsilon_{0}>0$
(see Fig. 2a), the magnetic boundary crosses the line $y=0$, and
the level magnetizes even when the impurity is above the Fermi energy.
This is understood by the fact that the hybridization leads to a large
broadening of the impurity level density of states (with a tail that
decays like $1/\omega$) that crosses the Fermi energy even when the
bare level energy is above it. In the opposite case of $\epsilon_{0}<0$
a similar effect occurs with the crossing of the magnetic boundary
along the $y=1$ line, something that also does not occur in ordinary
metals \cite{phillip}. This implies even when the energy of the
doubly occupied state is below the Fermi level ($\epsilon_{0}+U<\mu$),
because of the large broadening, the impurity magnetizes if $U$ is
not too large or too small. The up turn close to $y=1$ and $x\ll1$
in the $\epsilon_{0}<0$ case only reflects that in this limit ( $U$,$\,\mu\gg\epsilon_{0}$,
for finite $\epsilon_{0}$) the physics of the Dirac points is irrelevant
and we recover the usual Anderson model in ordinary metals, where
the transition curve approaches the point $x=0$, $y=1$ from below\cite{anderson61}.
This picture is physically consistent with a renormalization group
calculation at fixed $\mu=0$ \cite{fritz}.

The dependence of the scaling of the magnetic boundary with $\epsilon_{0}$
and $\Delta$ (see Fig. 2) shows that the size of the magnetic region
grows as $|\epsilon_{0}|$ approaches the energy of the Dirac points.
In this situation the DOS around the localized level is suppressed,
favoring the formation of a local magnetic moment. In particular,
in the limit where the level is nearly at the Dirac point ($|\epsilon_{0}|\to0)$,
the level nearly decouples from the bath and the impurity can magnetize
in principle for any small finite charging energy $U$. On the other
hand, the magnetic region shrinks in the $y$ direction as the hybridization
parameter $\Delta$ grows (see Fig.2). In the limit of $\Delta\to0$
and $U$ finite a local magnetic moment forms whenever $0<y<1$, as
in the case of an impurity in a metal.

\begin{figure}[t]

\begin{centering}
\includegraphics[clip,width=8.7cm]{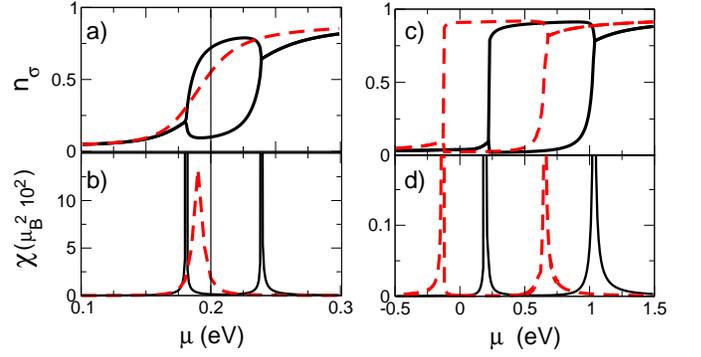} 
\par\end{centering}

\caption{ $n_{\uparrow}(\mu)$ ,$n_{\downarrow}(\mu)$ and $\chi(\mu)$ for
$|\epsilon_{0}|/D=0.029$ and $V/D=0.14$ ( $D\sim7$ eV). Left panels:
$x=11$ (dashed curves), and $x=5$ (solid). The impurity magnetizes
inside the bubble ($n_{\uparrow}\neq n_{\downarrow})$. The vertical
line marks the position of the level, $\epsilon_{0}=0.2$ eV. On the
right: $\epsilon_{0}=0.2$ eV (solid) and $\epsilon_{0}=-0.2$ eV
(dashed) at $x=0.45$.}

\label{sus} 
\end{figure}

The application of a potential $V_{g}$ through an electric field
via a back gate \cite{novo1} shifts the chemical potential $\mu$
and moves the magnetic state of the impurity in the vertical direction
($y$) in Fig.~2. We assume that $x=D\Delta/U$ does not change much
with applied voltage, even with screening coming from a finite Fermi
energy. Hence, the magnetization of the impurity can in principle
be turned on and off, depending only on the gate voltage applied to
graphene. This is better illustrated by looking at the behavior of
the impurity magnetic susceptibility. In the presence of a field,
the energy of the impurity spin states changes to $\epsilon_{\sigma}=\epsilon_{0}-\sigma\mu_{B}B+Un_{-\sigma}$.
In the zero field limit, the magnetic susceptibility of the impurity,
$\chi=\mu_{B}\sum_{\sigma}\sigma\left(\mbox{d}n_{\sigma}/\mbox{d}B\right)_{B=0}$
($\mu_{B}$ is the Bohr magneton, and $B$ is an applied magnetic
field), can be calculated straightforwardly from Eq. (\ref{MF}):
\begin{eqnarray}
\chi=-\mu_{B}^{2}\sum_{\sigma=\uparrow\downarrow}\frac{dn_{\sigma}}{d\epsilon_{\sigma}}\cdot\frac{1-U\frac{dn_{-\sigma}}{d\epsilon_{-\sigma}}}{1-U^{2}\frac{dn_{-\sigma}}{d\epsilon_{-\sigma}}\frac{dn_{\sigma}}{d\epsilon_{\sigma}}}\,.\label{finalsus}\end{eqnarray}
 In the lower panels of Fig. 3 we show $\chi(\mu)$ for $\epsilon_{0}=0.2$
eV, $V\sim1$ eV and $D\sim7$ eV, for different values of $x$. The
corresponding magnetization line for this set of parameters is defined
by the solid curve with black circles in Fig. 2a. While the impurity
remains non-magnetic for any $y$ at $x=11$ ($U\sim40$ meV), as
shown in Fig. 2a, the impurity state already crosses the magnetic
boundary twice for $x=5$, where $U$ is nearly twice larger. In this
case, a large magnetic moment of $\sim0.5$ $\mu_{B}$ forms below
the energy of the level, at $\mu\sim0.18$ eV (Fig. 3a and 3b). At
$x=0.45$ ($U=1$ eV), the local moment exists for very large $\mu\sim1$
eV, and a strong and uniform magnetic moment of $\sim0.9\,\mu_{B}$
forms in almost the whole magnetic region (see Fig. 3c). A similar
qualitative behavior for the magnetization occurs when $\epsilon_{0}<0$
(Fig. 3c,d). As $U$ becomes large ($>1$ eV), the magnetic transition
becomes very sharp. For $\epsilon_{0}\sim0.5$ eV and $V=1$ eV, the
impurity can magnetize for $U\gtrsim0.1$ eV. While the local charging
energy $U$ for transition metals in a metallic matrix is of the order
of $\sim$ 5$-$10 eV \cite{mahan}, in graphene, where the effective
hybridization can be large due to the linear increase of the DOS,
the critical $U$ for magnetization of the impurity can be much smaller.
Hence, transition elements and molecules that usually do not magnetize
when introduced in ordinary metals, can actually become magnetic in
graphene \cite{Duffy,Leenaerts}.

\begin{figure}[t]

\begin{centering}
\includegraphics[clip,width=8.65cm]{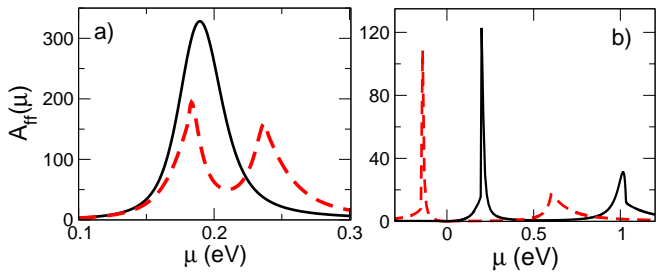} 
\par\end{centering}

\caption{Spectral function (in units of 1/eV) of the $f$-electrons at the
Fermi energy $\mu$ for $|\epsilon_{0}|/D=0.029$ and $V/D=0.14$
($D\sim7$ eV). (a) $x=11$ (solid curve) and $x=5$ (dashed) for
$\epsilon_{0}>0$ (see Fig. 2). (b) $x=0.45$ for $\epsilon_{0}>0$
(solid) and $\epsilon_{0}<0$ (dashed). }

\end{figure}

In order to show that the spectroscopic functions of the magnetic
impurities can also be controlled by electric field effect we show,
in Fig. 4, the spectral function of the localized electrons calculated
at the Fermi energy: $A_{ff}(\omega=\mu)=2\pi\sum_{\sigma}\rho_{ff,\sigma}(\mu)$.
The solid line in Fig. 4a is a non-magnetic resonance in a situation
where the impurity state does not cross the magnetic boundary of the
scaling diagram by changing $y$ ($\mu$) for some fixed $x$. In
the other curves of Fig.4, the spectral weigth splits between two peaks
located around the magnetic transitions, near $\mu\sim\epsilon_0$ and $\epsilon_0+U$ (see Fig.3).

The dependence of the impurity density of states with $\mu$, and
hence with gate bias, allows for the identification of the formation
of local moments through ordinary transport measurements. For finite
$\mu$, the hybridization of the itinerant electrons with the localized
level renormalizes the charge scattering channels and hence the carrier
conductivity, $\sigma=2e^{2}|\mu|\tau$, where $\tau^{-1}$ is the
impurity scattering rate. Second order perturbation theory gives \cite{mahan}:
\begin{equation}
\tau^{-1}-\tau_{0}^{-1}\propto n_{0}V^{2}A_{ff}(\mu)\,,\label{sigma}\end{equation}
 where $\tau_{0}^{-1}$ is the scattering rate of the electrons in
the absence of impurities and $n_{0}$ is the impurity concentration.
In the limit of very large $U$, however, the scattering is dominated
by the spin flip channels in the Kondo regime\cite{fradkin90,lu01,paco,dora07,Baskaran07}.
When $\epsilon_0$ is located in the experimental range accessible 
by the application of a gate voltage $\sim-0.3$ to 0.3
eV \cite{novo1}, the shape of the dip in the conductivity produced
by the impurity scattering can indicate not only the position of the
energy level but also the presence of local magnetic moments 
(notice that the non-magnetic resonance in the spectral function is quite symmetric).

In the presence of a finite density of magnetic moments a macroscopic
magnetic state can egress due to the RKKY interaction between them.
At the Dirac point ($\mu=0$) the interaction is purely ferromagnetic
due to the vanishing of the Fermi wavevector, $k_{F}=\mu/v_{F}$ \cite{maria}.
However, at finite bias voltage the RKKY interactions display $2k_{F}$
oscillations decaying like $1/r^{3}$ \cite{falko} that can couple
the magnetic moments ferromagnetically or anti-ferromagnetically depending
on the position and geometry of the adatom lattice (that can be conveniently
chosen using a STM). Hence, by changing the bias voltage a variety
of different magnetic phases can emerge.

In conclusion, we have examined the conditions under which a
transition metal adatom on graphene can form a local magnetic moment. We find
that due to the anomalous broadening of the adatom local electronic
states, moment formation is much easier in graphene. Furthermore,
the magnetic properties of adatoms can be controlled by electric field
effect allowing for the possibility of using graphene in spintronics.
We thank V. M. Pereira, J. M. B. Lopes dos Santos, A. Polkovnikov,
and S. W. Tsai for helpful discussions. NMRP acknowledge the financial
support from POCI 2010 via project PTDC/FIS/64404/2006. BU acknowledges
CNPq, Brazil, for the support under the grant 201007/2005-3.

\end{document}